\documentclass[10pt,a4paper,twocolumn,english,prl,aps,showpacs,floatfix,groupedaddress,superscriptaddress]{revtex4}
\usepackage[T1]{fontenc}
\usepackage[latin9]{inputenc}
\usepackage{amsmath}
\usepackage{amssymb}
\usepackage{braket}
\usepackage{esint}

\makeatletter

\pdfpageheight\paperheight
\pdfpagewidth\paperwidth

\providecommand{\tabularnewline}{\\}

\@ifundefined{textcolor}{}
{%
 \definecolor{BLACK}{gray}{0}
 \definecolor{WHITE}{gray}{1}
 \definecolor{RED}{rgb}{1,0,0}
 \definecolor{GREEN}{rgb}{0,1,0}
 \definecolor{BLUE}{rgb}{0,0,1}
 \definecolor{CYAN}{cmyk}{1,0,0,0}
 \definecolor{MAGENTA}{cmyk}{0,1,0,0}
 \definecolor{YELLOW}{cmyk}{0,0,1,0}
}

\@ifundefined{definecolor}
 {\usepackage{color}}{}
\@ifundefined{definecolor}{\@ifundefined{definecolor}
 {\usepackage{color}}{}
}{}\@ifundefined{definecolor}{\@ifundefined{definecolor}{\@ifundefined{definecolor}
 {\usepackage{color}}{}
}{}}{}\def\1{1\negthickspace{\rm I}}

\makeatother

\usepackage{babel}

\makeatother

\usepackage{babel}
\begin{document}

\title{Symmetry protected topological phases of 1D interacting fermions
with spin-charge separation}

\author{Arianna Montorsi}

\email{arianna.montorsi@polito.it}
\affiliation{Institute for condensed matter physics and complex systems, DISAT,
Politecnico di Torino, I-10129, Italy}

\author{Fabrizio Dolcini}
\affiliation{Institute for condensed matter physics and complex systems, DISAT,
Politecnico di Torino, I-10129, Italy}
\affiliation{CNR-SPIN, Monte S.Angelo - via Cinthia, I-80126 Napoli, Italy}

\author{Rita C. Iotti}

\affiliation{Institute for condensed matter physics and complex systems, DISAT,
Politecnico di Torino, I-10129, Italy}
\author{Fausto Rossi}
\affiliation{Institute for condensed matter physics and complex systems, DISAT,
Politecnico di Torino, I-10129, Italy}

\date{30 May 2017}
\begin{abstract}
The low energy behavior of a huge variety of 1D interacting spinful fermionic systems exhibits spin-charge separation, described in the continuum limit by two sine-Gordon models decoupled in the charge and spin channels. Interaction is known to induce, besides the gapless Luttinger liquid phase, eight possible gapped phases, among which the Mott, Haldane, charge-/spin-density,  and bond-ordered wave insulators, and the Luther Emery liquid. Here  we prove that some of these physically distinct phases have non-trivial topological properties, notably the presence of degenerate protected edge modes with fractionalized charge/spin. Moreover, we show that the eight phases are in one-to-one correspondence with the symmetry protected topological (SPT) phases classified by group cohomology theory in the presence of particle-hole symmetry P. The latter result is also exploited to characterize SPT phases by measurable non-local order parameters which follow the system evolution to the quantum phase transition. The implications on the appearance of exotic orders in the class of microscopic Hubbard Hamiltonians, possibly without P symmetry at higher energies, are discussed.
\end{abstract}


\maketitle

\section{Introduction}
\noindent
Interacting one-dimensional (1D) systems are known to display peculiar features that cannot be accounted for by Fermi-liquid theory.
A wide class of physically relevant 1D models of interacting fermions, including the Hubbard model~\cite{Hubb}, exhibits an effect known as spin-charge separation: at low energies, charge and spin behave as two independent degrees of freedom, whose dynamics is well described  within the bosonization formalism by two decoupled sine-Gordon models~\cite{Giam}. 
A renormalization group (RG) analysis shows that interaction can lead to 9 possible phases~\cite{GNT}: the gapless phase is the Luttinger liquid (LL), whereas the remaining 8 phases do have a  gap in either the charge or the spin channel (partly gapped phases), or in both (fully gapped phases)~\cite{JaMH,JaKa,DoMo}. These gapped phases are the Luther-Emery (LE) liquid, Haldane (HI) and Mott insulators (MI), and variuos types of density wave (DW) insulators. To each of the gapped phases a non-vanishing nonlocal local order (NLO)  parameter\cite{MoRo} can be associated, in some cases of Haldane type\cite{BMR}. Notably, the low-energy phase diagram is derived from the asymptotic properties of the sine-Gordon model, and does not specifically rely on any topological argument. However, it was successively understood that one of the LE liquid phases exhibits in fact topological features in specific cases \cite{KaCa, KeBe}. 

The interesting open question is thus whether in general the above phases identfied by RG analysis may host non-trivial topological properties, and eventually be classified {\em  topologically}. Because the gaps of these phases are purely induced by interaction, rather than simply modified by it, the standard topological classification adopted for non-interacting systems~\cite{AZ,SRFL} does not apply, and different approaches  must be invoked. 
On general grounds, the presence of non-local orders (NLO) in low temperature phases of
quantum matter \cite{Niro,Eal}, and its intertwining with topological
orders\cite{Wen,NuOr} did lead in the last decades to a deeper understanding
of the possible collective behaviors of quantum systems in ordered
phases which do not break continuous Hamiltonian symmetries. 
In particular, the concept of symmetry protected topological (SPT) phases, first introduced
for spin systems \cite{PTBO,GuWe,PBTO}, allows one to describe how
interaction modifies the topological classification of gapped non interacting
insulators and superconductors\cite{AZ,SRFL} in presence of discrete Hamiltonian symmetries. The phases are denoted as symmetry protected, since they are robust with respect to perturbations of the Hamiltonian which preserve its symmetries: the phases with non trivial topological properties, such as the localization of degenerate edge modes, will remain distinct from the trivial ones until the system undergoes a quantum phase transition. 
\\In fact, the concept of SPT phases turns out to be far more general, providing an exhaustive classification of the distinct ordered phases of interacting bosonic and fermionic lattice systems \cite{CGW,CGLW,CGLW2} (see also \cite{Wen-1} for a comprehensive review). In one dimension, it has also been well understood in terms of entanglement properties of their microscopic degrees of freedom\cite{PTBO,SPC}.  In particular, the approach of SPT phases could be applied 
\cite{Fuji} to discuss the topological classification of  gapped phases of systems which are gapless in the non-interacting limit, like the 1D interacting fermion models mentioned above.  

This paper is devoted to determine the interaction induced topological properties of the low temperature phase diagram of 1D fermions with  spin-charge separation.  We show that the low-energy sector reveals symmetries that, despite being broken at higher energies, characterize physically interesting regimes and encode topologically useful information. In particular, by exploiting the discrete P symmetry and the continuous U(1) symmetries of the two decoupled sine- Gordon models, we prove that the 8 gapped phases deduced by RG analysis are in one-to-one correspondence with the cohomology group classification of SPT phases. This result bridges the  ground state phase diagram of these systems --consisting of MI, HI, two LE liquids, and 4 distinct DW insulating phases-- with the SPT classification. It thus enables us to predict non-trivial topological effects, such as the existence of edge states in some of the above phases, on the basis of the presence of the appropriate symmetry in the low energy sine-Gordon model. As a further byproduct, our approach also allows one to associate to distinct SPT phases appropriate NLO parameters. 

The paper is organized as follows. In Sec.II we present the decoupled sine-Gordon model for spin-charge separation, motivating its adoption as the low energy sector of a wide class of 1D microscopic fermionic models.
In Sec.III we summarize the low energy phases obtained from the RG analysis of the model, and discuss their NLO. In Sec.IV we present our results concerning the topological classification of these phases. We first exploit the symmetries emerging at low energies to identify the symmetries protection acting on each phase of  the RG phase diagram. Then, we show that some phases host protected edge modes. Finally, by applying the cohomology group theory we determine the topological classification of the full phase diagram. In Sec. V we discuss our results, and outline possible future developments.

\section{Model Hamiltonian in the low energy limit}

It is well known that the low-energy physics of a wide class of 1D interacting fermion systems at half band filling is characterized by spin-charge separation. This is well captured by an effective Hamiltonian  consisting of two decoupled sine-Gordon models  describing the charge and spin sector, respectively, i.e.\\
\begin{equation}
{\cal H}_{SG}=\sum_{\nu=c,s}\int dxH_{\nu}(x)\label{eq:Ham}
\end{equation}
with

\begin{equation}
H_{\nu}=H_{0\nu}+\frac{m_{\nu}v_{\nu}}{2\pi a^{2}}\cos(\sqrt{8\pi}\phi_{\nu})\label{eq:Ham_SG}
\end{equation}
and 

\begin{equation}\label{eq:H0nu}
H_{0\nu}=\frac{v_\nu}{2}\Big[K_{\nu}(\partial_{x}\theta_{\nu})^{2}+\frac{1}{K_{\nu}}(\partial_{x}\phi_{\nu})^{2}\Bigr]\quad.
\end{equation}
Here $\Pi_\nu=-\partial_x \theta_\nu$ is the conjugate momentum of the field $\phi_\nu$. The mass coefficients $m_\nu$, the Luttinger parameters $K_\nu$ and the velocities $v_\nu$ depend on the coupling constants and the filling of the original microscopic lattice Hamiltonian\cite{JaKa,AlAr,Naka,Fabr,DoMo}. 
Each sector $\nu=c,s$ of the model (1) is characterized by a competition between the $H_{0\nu}$ term (\ref{eq:H0nu}), which describes an harmonic oscillator with fluctuating field $\phi_\nu$, and the cosine term, which would favor the pinning of the field around a locked value. Such competition leads to the appearance of various physical phases, which we shall summarize in the next section. \\

A comment is now in order. Later in the paper we shall also refer to lattice models and quantities, formulated in terms of fermionic operators $c_{j\sigma}$, with $j$ denoting the lattice site and $\sigma=\pm 1$ the spin projection. It is thus worth recalling here briefly the connection between the continuum model (\ref{eq:Ham})-(\ref{eq:Ham_SG}) and the underlying microscopic lattice models. The idea is that  in most relevant situations  the physically interesting energy range of a lattice model is small as compared to the whole bandwidth. It is thus customary to introduce an effective model which describes the low-energy sector. To this purpose, one first rewrites \cite{JaMH} the original lattice Hamiltonian in terms of four fermionic fields $\psi_{\alpha \sigma}(x)$ on the continuum, with $\alpha=\pm 1$ referring to the two Fermi points $\pm k_F$, related to the lattice operators via $\psi_{\alpha \sigma}(x) = \sum_{k \simeq \alpha k_F} e^{i k x} c_{k \sigma}$, where $c_{k \sigma}$ are the Fourier components of the original lattice operators $c_{j \sigma}$.
Then, by adopting the bosonization formalism\cite{Giam,GNT}, one re-expresses the four fermionic fields as vertex operators $\psi_{\alpha \sigma}= \eta_{\alpha \sigma}  \exp \left[i \sqrt{\frac{\pi}{2}}(\alpha \phi_c+\alpha \sigma \phi_s +\theta_c +\sigma \theta_s)\right]/{\sqrt{2\pi a}}$, where $\phi_\nu,\theta_\nu$ (with $\nu=c,s$) is a pair of bosonic fields related to charge (c) or spin (s) degrees of freedom, $a$~is a short-distance cut-off, and $\eta_{\alpha\sigma}$ are anticommuting Klein factors. 
By applying  this method e.g. to the Hubbard model~\cite{Hubb}, which describes the competition of a band "hopping" term and an "on-site" 2-body Coulomb repulsion and which has found a broad spectrum of applications in both hard and soft condensed matter physics, the model (\ref{eq:Ham})-(\ref{eq:Ham_SG}) is obtained. The same occurs for a wide class of  extended Hubbard-like models that include further one-body terms (such as spin-orbit coupling or next-nearest neighbors hopping), two-body interaction beyond the on-site approximation (such nearest neighbor diagonal interaction, correlated hopping, pair hopping and so on)~\cite{Hubb}, as well as 3- and 4-body interactions\cite{DoMo1} that may be relevant in ultracold atoms or molecules trapped in optical lattices.  

For these reasons, the sine-Gordon model (\ref{eq:Ham})-(\ref{eq:Ham_SG}) plays a crucial role in the characterisation of 1D fermion models and is the starting point of our analysis.

\section{Ground state phase diagram and non-local order parameters}

The result of the competition of the two terms appearing in Eq.(\ref{eq:Ham_SG}) depends on the sign and the magnitude of the mass term $m_\nu$ in each channel $\nu$, and can be established by a well known RG analysis of the sine-Gordon model\cite{GNT}. 
Let us briefly recall the main results:  for $4(\sqrt{K_\nu}-1) < |m_\nu|$, the cosine term becomes relevant,  and a gap opens in the related channel $\nu=c,s$, with the field $\phi_\nu$ pinned around either 0 (for $m_\nu<0$) or $\sqrt{\pi/8}$ (for $m_\nu>0$). The case where no gap opens in either channel corresponds to the Luttinger liquid (LL) phase, where both $\phi_c$ and $\phi_s$ are unpinned and the effective Hamiltonian is $H_{LL}=\sum_\nu \int H_{0\nu}$. The remaining 8 phases, characterized by a gap in at least one channel, are listed on the left-hand side of  Table~I, with the pinning value of the fields provided in the first and second column.

In the Table, the LE and Haldane LE (HLE) liquid phases, and the MI and HI insulating phases are partly gapped, i.e. characterized by the presence of a gap in only one channel, namely the spin or the charge respectively. In contrast,  BOW, CDW, SDW and BSDW  are fully gapped phases, also characterized by a spontaneous symmetry breaking. 

The explicit expression of the parameters in ${\cal H}_{SG}$ in terms of the coupling constants of the microscopic Hubbard-like models were given in previous works \cite{JaKa,AlAr,Naka,Fabr,DoMo} and will be not reported here.  We  emphasize that all the phases reported on the left-hand side of  Table~I can be reached for appropriate values of the microscopic parameters, the only exception being  the HLE phase: within bosonization, this is reached for instance by adding a further spin-orbit coupling term~\cite{GJPB, KeBe, KaCa} (see also section \ref{symm}).

\begin{table}
\begin{tabular}{cc}

\begin{tabular}{|c|c|c|c|}
\hline 
 & $\sqrt{8\pi}\phi_{c}$ & $\sqrt{8\pi}\phi_{s}$ & NLO  \tabularnewline
\hline 
\hline 
{\footnotesize  LE}  & $u$ & 0 & ${\cal O}_{\cal P}^{s}$   \tabularnewline
\hline 
{\footnotesize  MI}  & 0 & $u$ & ${\cal O}_{\cal P}^{c}$  \tabularnewline
\hline 
{\footnotesize HLE} & $u$ & $\pm\pi$& ${\cal O}_{\cal S}^{c}$   \tabularnewline
\hline 
{\footnotesize HI} & $\pm\pi$ & $u$ & ${\cal O}_{\cal S}^{c}$  \tabularnewline
\hline 
{\footnotesize BOW} & 0 & 0  & ${\cal O}_{\cal P}^{c},\:{\cal O}_{\cal P}^{s}$ \tabularnewline
\hline 
{\footnotesize CDW} & $\pm\pi$ & 0 & ${\cal O}_{\cal S}^{c},\,{\cal O}_{\cal P}^{s}$  \tabularnewline
\hline 
{\footnotesize SDW} & 0 & $\pm\pi$ & ${\cal O}_{\cal P}^{c},\:{\cal O}_{\cal S}^{s}$ \tabularnewline
\hline 
{\footnotesize BSDW} & $\pm\pi$ & $\pm\pi$ & ${\cal O}_{\cal S}^{c},\:{\cal O}_{\cal S}^{s}$ \tabularnewline
\hline 
\end{tabular}

\begin{tabular}{|c|c|}
\hline 
  {SP} & {GCC} \tabularnewline
\hline 
\hline 
 triv & $\lambda_{s}=1$  \tabularnewline
\hline 
  triv & $\lambda_{c}=1$ \tabularnewline
\hline 
  P,T& $\lambda_{s}=-1$ \tabularnewline
\hline 
 P & $\lambda_{c}=-1$ \tabularnewline
\hline 
  triv & $\lambda_{s}=1=\lambda_{c}$\tabularnewline
\hline 
  P & $\lambda_{s}=1=-\lambda_{c}$ \tabularnewline
\hline 
  P,T & $\lambda_{c}=1=-\lambda_{s}$ \tabularnewline
\hline 
  P & $\lambda_{s}=-1=\lambda_{c}$ \tabularnewline
\hline 
\end{tabular}

\end{tabular}

\caption{
Left-hand side: Classification of massive phases for ${\cal H}_{SG}$. The first two columns denote the pinning values for the charge (c) and spin (s) fields $\phi_\nu(x)$ ($\nu=c,s$), respectively, with $u$ staying for "unpinned", while the third column  denotes the non vanishing NLO parameters ${\cal O}_{\cal A}^{\nu}$. \\Right-hand side: the topological characterization of the phases obtained in Section IV. In the fourth column the RG phase without topological features are denoted by  "triv", whereas we report the type of symmetry (T and or P) protecting the RG phases with degenerate edge modes. The last column describes the trivial ($\lambda_{\nu}=+1$) and non trivial ($\lambda_{\nu}=-1$) SPT phases obtained within the group cohomology classification (GCC).}
\end{table}

\paragraph{Non local order parameters.}
The standard way of characterizing the physical features of 
different ordered phases is to analyze  the asymptotic behavior
of correlation functions of appropriate local operators. In 1D,
 they asymptotically decay to zero  with a power or exponential
law: the ``order'' of the phase is described by the correlation
function which decays slower. In case of fully gapped phases with
spontaneously broken symmetries, one of the correlation functions
does remain finite, thus capturing the long range order of the phase through a local operator,
 whose expectation value can be regarded to as an order parameter.\\
Recently it has been realized \cite{BDGA,MoRo,BMR} that, even when none
of the Hamiltonian symmetries is broken (as in a partly gapped phase),
 order parameters can be identified with appropriate operators, which average value is non-zero 
in these phases and vanishes  at the phase transition. Importantly, such operators have to be non-local
in the lattice representation. They can be built by means of the on site charge and spin operators  $s_{\nu j}$ as:
$s_{cj}=(n_{j\uparrow}+n_{j\downarrow}-1)/2$, and $s_{sj}=(n_{j\uparrow}-n_{j\downarrow})/2$,
$n_{j\sigma}\doteq c_{j\sigma}^{\dagger}c_{j\sigma}$ being the fermion
number operator at site $j$ with spin $\sigma.$ The NLO operators can now be expressed as  
${\cal O}_{\cal P}^{\nu}(r)=\exp [2\pi i\sum_{j=1}^{r-1} s_{\nu j} ]$,
and ${\cal O}_{\cal S}^{\nu}(r)=\exp [2\pi i\sum_{j=1}^{r-1}s_{\nu j} ](2s_{\nu r})$. 
 In the continuum limit, the argument of the exponential acquires the  form  
\begin{equation}
S_\nu^{z}(r)=\int^r dx \rho_{\nu}(x)\quad , \label{eq:snu}
\end{equation} 
where the densities $\rho_{\nu}(x)$ are related to the bosonic fields through: $\rho_{\nu}(x)= \partial_{x}\phi_{\nu}/\sqrt{2\pi}$. Then ${\cal O}_{\cal A}^{\nu}(r)$, with ${\cal A =P,S}$ are obtained respectively as the symmetric
and anti-symmetric form of $\exp[i2\pi S_{\nu}^z (r)]$~\cite{BDGA}.
Explicitly, 
\begin{equation}
{\cal O}_{\cal P}^{\nu}\thicksim\cos(\sqrt{2\pi}\phi_{\nu})\quad,\quad{\cal O}_{\cal S}^{\nu}\sim\sin(\sqrt{2\pi}\phi_{\nu})\quad,\label{eq:strings}
\end{equation}
generating the four asymptotic correlation functions
\begin{equation}
{\cal C}_{\cal A}^{\nu}=\lim_{r\rightarrow\infty}\langle{\cal O}_{\cal A}^{\nu\dagger}(0)\ {\cal O}_{\cal A}^{\nu}(r)\rangle\quad.\label{eq:string}
\end{equation}

In each channel, these are also known as parity and Haldane string correlators respectively.
They all vanish when the fields are unpinned (LL phase). However,
when at least one of the fields $\phi_{\nu}$ is pinned to one of the
two values reported in  the left-hand side of  Table~I, in each channel one of the correlation functions
${\cal C}_{\cal P}^{\nu}\propto\langle\cos\sqrt{2\pi}\phi_{\nu}\rangle^{2}$, ${\cal C}_{\cal S}^{\nu}\propto\langle\sin\sqrt{2\pi}\phi_{\nu}\rangle^{2}$
remains finite\cite{MoRo}  identifying a long-range NLO, whereas the
other one still vanishes. The average values of  ${\cal O}_{\cal A}^{\nu}$ (${\cal A=P,S}$)
are thus identified as order parameters in the $\nu$ channel for the
two possible gapped phases. In this way the bosonization description
is able to associate appropriate
NLO parameters to each partly or fully gapped phase, as arized in the last column of the left-hand side of Table~I. In particular, in the Table we denoted the phases characterized by  only one non-vanishing Haldane string correlator, namely ${\cal A=S}$ 
in the $\nu$ channel, as HLE ($\nu=s$) and HI ($\nu=c$) respectively.\\

The discrete lattice expressions for $s_{\nu j}$, when inserted 
into the non local operators ${\cal O}_{\cal A}^{\nu}$, clarify the different microscopic arrangements within 
the two partly gapped phases which may take place in each channel. In fact,  
the charge degrees of freedom may be identified with the local configurations with $s_{cj}\neq0$, i.e.
empty sites (holons) and doubly occupied sites (doublons). Whereas the spin degrees of freedom are
those configurations for which $s_{sj}\neq0$, i.e., sites occupied by a single fermion with either up or down spin.
The finiteness of the Haldane string correlator (\ref{eq:string})
implies that for $\nu=c$ in the HI phas  holons and doublons are alternated (i.e. between two successive doublons there is always a holon) and intercalated by an arbitrary number of single fermions. In contrast, for $\nu=s$ the Haldane LE liquid phase is metallic: the alternated (up and down) single spins dilute
in the background of holons and doublons. In the latter case the results here
complement and generalize the findings of Refs.[\onlinecite{KeBe,KaCa}], 
where it was shown that the addition of spin-orbit coupling to a metal
may generate topological effects: this will happen when the string Haldane NLO becomes non-vanishing.  The parent ground states for these phases are reached in the strong coupling lmit when the string order parameter 
saturate to the value 1; in this case $\nu$ degrees of freedom form singlets on neghboring sites. 
\\A non-vanishing value of the parity operators denotes instead a state which  --in the strong coupling classical limit  where ${\cal C}_{\cal P}^{\nu}=1$ 
(parent ground states)-- is a direct product of on-site $\bar{\nu}$
states, and is highly degenerate. For $\nu=c$, it describes an insulator
of strictly singly occupied sites, as in the case of the infinite-$U$
repulsive Hubbard model, whereas for $\nu=s$  the parent ground state
consists solely of holons and doublons, as in the case
of the attractive infinite-$U$ Hubbard model. In both cases at finite
coupling  strength  entangled parity breaking pairs appear, removing the degeneracy of the $\bar{\nu}$
states.

\section{Classification of symmetry protected topological phases}

The various phases described in the previous section and summarized on the left-hand side of the Table~I are characterized in terms of their gaps, field pinning values, and NLO parameters. We emphasize that the well known phase diagram is purely obtained from the RG analysis of the asymptotic behavior of sine-Gordon model, without invoking any topological argument. In this section we shall now provide the topological characterization of the phases.  As mentioned above, topological features may emerge under appropriate symmetry protection.  

 We start by recalling that the tenfold way classification of distinct topological phases,
proposed by Altland and Zirnbauer\cite{AZ,SRFL} for non-interacting  gapped systems described by a bilinear fermionic Hamiltonian with some symmetry, becomes unstable in presence of two-body interaction \cite{FK,MFM}. 
The role of instability is the reduction of the non-interacting
classification to a finite number of phases in each class, which can be identified depending
on the group $G$ of symmetries of the Hamiltonian and of the symmetry group $G'\subseteq G$
of the ground state. Such phases are distinct under
the symmetry protection of $G'$ as long as they cannot be reduced
to the trivial one by any adiabatic transformation which preserves
$G'$. Given the symmetry groups of both the Hamiltonian and its ground
state, the SPT phases classification for interacting systems holds
in the strong coupling limit, providing a general classification
of the distinct phases of these systems. In particular, for 1D systems  
an extremely powerful SPT phases classification method is the so called group 
cohomology theory\cite{CGW}, based on the inspection of the local projective 
representations of $G'$ on the low-energy edge states. This classification, which is also closely related to the the entanglement properties on the two ending sites of a chain bipartition\cite{PBTO,TPB}, predicts the existence of 
phases that are non-trivial from the topological point of view, the
most striking feature being the existence of degenerate protected
edge modes. Distinct phases can be recognized
also in absence of topological properties, and are known as trivial SPT 
phases.   

Our strategy is based on the observation that the non interacting fermionic case may as well describe a gapless system with no topological properties, at variance with the non interacting topological superconductor/insulator. In this case it is the interaction which could drive the system to open gaps. This happens for instance for the sine-Gordon model  ${\cal H}_{SG}$, whose non-interacting limit describes free massless Dirac fermions\cite{Giam,GNT}. 
Hence we first exploit the symmetries of ${\cal H}_{SG}$  to  prove that all the massive phases are distinct under symmetry protection. Then, we identify the presence of degenerate symmetry protected edge modes in the phases characterized by non-vanishing Haldane NLO. Finally, we apply the group cohomology theory to classify the SPT phases in the strong coupling limit, establishing a one-to-one correspondence between the gapped phases identified by RG classification (left-hand side of  Table~I), and their topological classification obtained by means of group cohomology.  These results, summarized on the right-hand side of  Table I, enable us to distinguish on general grounds  topologically trivial and non-trival phases in the RG classification, and in turn to associate a NLO parameter to each of the topological phases of group cohomology classification (GCC). Here below the detailed analysis is presented.

\subsection{Symmetries}\label{symm} The Hamiltonian ${\cal H}_{SG}$ has both time-reversal T and particle-hole
P discrete symmetries. {Indeed, since in real space $T\psi_{\alpha\sigma}T^{-1}=\sigma\psi_{\bar\alpha \bar\sigma}$ and 
$P\psi_{\alpha\sigma}P^{-1}=\psi^\dagger_{\alpha \sigma}$, one can derive from the bosonized expression of the fermionic fields given above the action of both symmetries on the bosonic fields $\phi_{\nu}(x)$, $\theta_{\nu}(x)$} as
\begin{equation}
\begin{aligned}T\phi_{\nu}T^{-1}=\delta_{\nu}\phi_{\nu}\;,\;T\theta_{\nu}T^{-1}=-\delta_{\nu}\theta_{\nu}\end{aligned}
,\label{eq:tr}
\end{equation}
\begin{equation}
P\phi_{\nu}P^{-1}=-\phi_{\nu}\;,\;P\theta_{\nu}P^{-1}=-\theta_{\nu} \; ,\label{eq:ph}
\end{equation}
{with $\delta_c=1$, $\delta_s=-1$. From the above relations} it is easily verified that T and P are symmetries of both $H_{\nu}(x)$s. \\It must be stressed that in fact, at variance with the lattice model which could possibly break these symmetries, their presence is a feature of the non-interacting spectrum linearized around the Fermi points
in the continuum limit, namely $v_F \sum k( n_{kR\sigma}- n_{kL\sigma})$.
Indeed, the action of T and P on right and left movers reads $T:c_{k\alpha\sigma}\rightarrow\sigma c_{-k\bar{\alpha}\bar{\sigma}}$
, and $P:c_{k\alpha\sigma}\rightarrow c_{-k\alpha\sigma}^{\dagger}$,
with $\bar{\sigma}=-\sigma$ ($P^{2}=1=-T^{2}$, $[T,P]=0$). Hence
our considerations apply to a larger class of models with respect
to those which exhibit these symmetries already on the lattice Hamiltonian.
Notice that for specific cases, like e.g. the Hubbard model itself, the invariance under T and P turns out 
to be a feature of the full spectrum, beyond the low-energy limit, with T: $c_{j\sigma}\rightarrow\sigma c_{j\bar{\sigma}}$,
and P: $c_{j\sigma}\rightarrow(-)^{j}c_{j\sigma}^{\dagger} $, the symmetries holding at zero-magnetization/half-filling respectively. {However, still within the class of Hubbard-like microscopic Hamiltonians, important cases such as those with correlated and/or next nearest neighbor hopping, and those with 3- and 4-body terms, do not exhibit P symmetry.} Nevertheless, the symmetries arise at low energies in the corresponding bosonized Hamiltonian ${\cal H}_{SG}$ \cite{Giam}. {Notice that P symmetry could hold also when right and left movers have different linearized dispersion relations around the Fermi points, as it may happen for instance in presence of a magnetic field.}

Apart from the above discrete $Z_{2}$ symmetries, ${\cal H}_{SG}$
turns out to have a U(1)$\times$U(1) continuous symmetry under $\theta_{\nu}\rightarrow\theta_{\nu}+const$.
Indeed both total charge and total spin, namely $S_{\nu}^z(L)$ --where $L$ is the length of the 1D system and $S_{\nu}^z(x)$ is defined in eq (\ref{eq:snu})-- are conserved quantities. For the underlying one-dimensional lattice
models these quantities read $S_{\nu}^{z}\doteq\sum_{j}s_{\nu j}$,
where $s_{\nu j}$ have been defined in terms of on-site fermonic operators in the previous section. Notice that the global symmetry can be realized locally, a fact which was both crucial in the construction of NLO operators in the previous section, and will be important in the implementation of the group cohomology classification in this section.  
\\In fact, in many cases the lattice Hamiltonians also have full rotational SU(2)
symmetry, which implies on the bosonized model that $K_{s}$ and $m_{s}$
cannot vary independently: in the weak coupling limit, a gapped spin
phase may appear only for $m_{s}<0$. This is the case for instance
for the Hubbard model. For the sake of generality, we release such
constraint so that the spin rotational symmetry reduces from SU(2)
to U(1), in principle allowing the dynamical opening of a spin
gap also for $m_{s}>0$. This may happen in presence of Rashba type
spin-orbit couplig\cite{GJPB,KeBe,KaCa}, which arises e.g. in InAs
wires\cite{Faal}, and can also be mimicked in optical lattices by
dephasing two one-dimensional lattices of spin 1/2 atoms in a transverse
magnetic field\cite{Citro}. 

Obviously, both charge and spin fermion parity, $\Pi_{\nu}=(-)^{2S_{\nu}^{z}}$,
are also preserved.

\subsection{Symmetry protection and edge modes}

Given the symmetries of the Hamiltonian, we now argue whether these are capable of protecting 
the robustness of the different phases with respect to symmetry preserving transformations. In fact, 
it was observed  \cite{AnRo} that in case of fermionic systems phases 
characterized by Haldane NLO may become fragile {(i.e. not distinct from the phase 
with parity NLO)} under appropriate choice of interaction. 
Such behavior originates from the presence --at any finite value of interaction-- 
of quantum fluctuations which may end up in adiabatically connecting the two phases\cite{MoPo}. 
This is not the case here. Indeed, the assumed spin-charge separation prevents by definition 
the possibility that quantum fluctuations in one (spin or charge) channel  connect phases which appear 
distinct in the other channel. Moreover in each channel the interaction, when relevant, induces the 
opening of a gap in two distinct ways, due to the pinning 
of the corresponding bosonic field $\sqrt{8\pi}\phi_{\nu}$ to one of the two values $0,\pi$. Hence the 
phases will remain distinct as far as it is not possible to change adiabatically one value into the other by a transformation.  
This is guaranteed by symmetry protection, since terms proportional to 
\begin{equation}
\sin(\sqrt{8\pi}\phi_{\nu})\label{eq: sin} \quad ,
\end{equation}
which would force the fields to pin to intermediate values, are not allowed. In particular, from Eqs. (\ref{eq:tr})-(\ref{eq:ph})
we see that P symmetry prevents such terms in the
charge channel, whereas both P and T symmetries protect the distinct
trivial and non-trivial phases in the spin channel. We thus realize
that the phases reported on the left-hand side of Table I are actually distinct SPT phases protected by T and/or P symmetries.

Having shown that the phases obtained within RG analysis are robust, we now discuss whether some of them
host protected edge modes. Upon generalizing
the argument proposed in \cite{KeBe,KaCa}, one sees that, when the
bosonic field in $\nu$ channel is pinned to the value $\pm\sqrt{\frac{\pi}{8}}$,
a kink $s_\nu$ of fractional charge/spin accumulates at the interface between
such phase and the trivial one ($\phi_{\nu}=0$). Explicitly, upon defining $s_\nu(x)=\lim_{a\rightarrow 0^+}\bigl [S_\nu^z(x)-S_\nu^z(x-a)\bigr ]$, with $S_\nu^z(x)$ given by (\ref{eq:snu}), one obtains
\begin{equation}
s_{\nu}(x)=\lim_{a\rightarrow 0^+}\frac{1}{\sqrt{2\pi}}[\phi_\nu(x)-\phi_\nu(x-a)]=\pm\frac{1}{4}\quad .\label{eq:edge}
\end{equation}
The latter identity holds at the edge between the two phases, where it is the hallmark of the presence of half the charge/spin of an electron. When the two configurations with different $s_{\nu}$ are degenerate
in energy --as for instance may happen for lattice models at half-filling
and zero magnetization-- two degenerate protected fractionalized edge modes are
thus realized, corresponding to $\phi_{\nu}=\pm\sqrt{\frac{\pi}{8}}$.\\ 
The above results are summarized in the fourth column of Table I, where the possible symmetries protecting each of the phases with fractionalized edge modes are indicated. The remaining phases are denoted as trivial.

 \subsection{Group Cohomology classification}
The one dimensional symmetry protected RG phases identified in the two previous subsections --with and without fractionalized edge modes-- can be put in one to one correspondence with the SPT phases classification one would obtain in the framework of group cohomology\cite{GuWe,CGW,CGLW,CGLW2}. In order to derive the latter, one should inspect the projective representations  of the symmetry group G of the Hamiltonian ${\cal H}_{SG}$ on its degenerate edge states.    As discussed in subsection IV.A, each  $\nu=c,s$ channel of the Hamiltonian is characterized by a continuous ${\rm U}(1)$ symmetry, preserving the total charge ($\nu=c$) and spin-$z$ component ($\nu=s$) operators. An element of the U(1) symmetry group can be rewritten as $U_\nu(\beta)=e^{i\beta_\nu 2S_\nu^z(L)}$ with $\beta_\nu \in [0,2\pi)$. Furthermore, the Hamiltonian of each channel also exhibits two discrete symmetries, T and P. From the right hand side of Table I, however, we notice that while P protects all phases, T guarantees protection within the spin channel only. We thus focus on P as discrete symmetry. The  symmetry group in each channel  is then ${\rm G} \equiv {\rm U}(1) \rtimes Z_2$, where the semidirect product is due to the fact that $S_\nu^z(L)$ and P do not commute. Indeed from Eqs.(\ref{eq:snu}) and (\ref{eq:ph}) one has 
\begin{equation}\label{eq:alg}
P\, U_\nu(\beta)=U_\nu(-\beta)\, P\quad.
\end{equation}
To apply the group cohomology classification, a further important requirement on G is that its elements  should be "local symmetries", meaning that the generators have a local representation at each $x$. For P this is ensured by the fact that its action on the fields $\phi_\nu$, given in Eq.  (\ref{eq:ph}), is in fact local. As for $U(\beta)$, we first notice that in the lattice representation it would read $U(\beta)=\prod_j U(\beta,j)$ with $U(\beta,j)={\rm e}^{i 2 \beta s_{\nu j}}$. In the continuum model, where $x=ja$ and $s_{\nu j}\rightarrow s_\nu(x)$ (see right identity in Eq.(\ref{eq:edge})), one has $U(\beta,x)=e^{i \beta 2 s_{\nu j}(x)}$. In particular, at the edge, $s_\nu =\frac{n}{2}$ for $\phi_\nu=2n\sqrt{\pi/8}$, and $s_\nu=(2m+1)/4$ for $\phi_\nu=(2 m+1)\sqrt{\pi/8}$, with $n,m\in {\bf Z}$. We can now inspect the projective representations $M$ of  U(1)$\rtimes Z_{2}$ on the two corresponding degenerate edge ($E$) states \cite{CGLW2, WaWe}. Explicitly, dropping the subscript $\nu$ and denoting $U_E(\beta) \equiv U(\beta,x=E)$, in each channel we may choose
\begin{equation} \label{eq:MU}
M[U_E(\beta)]=  {\rm e}^{i \frac{\beta}{2}n} \Biggl( \begin{matrix}
1&0\\0& {\rm e}^{i \frac{\beta}{2}m}
\end{matrix} \Biggr) \quad ,
\end{equation} 
and
\begin{equation} \label{eq:MP}
M(P)=\Biggl( \begin{matrix}
0&1\\ 1&0
\end{matrix} \Biggr)\quad ,
\end{equation}  
with $[M(P)]^2={ I_2}$. Other choices are  also possible. When inserting the above representation in (\ref{eq:alg}), the latter becomes 
\begin{equation}
M(P)M[U_E(\beta)]=  {\rm e}^{i\beta (n+\frac{m}{2}) }M[U_E(-\beta)]M(P) \quad , 
\end{equation}
which differs from (\ref{eq:alg}) for a phase factor. Only for even $m$ such phase factor can be gauged away upon redefinition of the overall phase in (\ref{eq:MU}), $\tilde M [U_E(\beta)]= {\rm e}^{i \frac{\beta}{2}\kappa}M[U_E(\beta)]$, with $\kappa\in {\bf Z}$.  Thus, depending on whether $m$ is even or odd, one identifies in each channel two inequivalent projective representations of  G, which we conventionally denote as  $\lambda_\nu=+1,-1$ respectively. The inequivalent projective representations of a given group are the elements of its second  cohomology group $H^2$(G,U(1)). The latter being a discrete group, it is not possible to connect continuously two of its elements: the corresponding phases are distinct. In particular, $\lambda_\nu=-1$ denotes the non-trivial projective representation, and corresponds to the non-trivial topological phase\cite{CGW, CGLW2}. 
\\In this way, the group cohomology analysis identifies in each channel, besides the gapless phase, two gapped phases, of which one topologically non-trivial. As a whole, 1 gapless and 8 distinct gapped SPT phases are predicted within such classification. We realize that these are in one to one correspondence with the symmetry protected spin and/or charge gapped phases identified by means of the RG analysis of 1D LL with sine-Gordon interaction. The correspondence is emphasized in the last column at r.h.s. of Table I. There, in each channel the non-trivial SPT phase obtained within group cohomology is naturally associated to the symmetry protected RG phase which hosts degenerate edge modes, according to the results obtained in the previous subsection. The correspondence can be further exploited:  each distinct trivial (non-trivial) SPT phase can be endowed with the appropriate parity (Haldane) string operator ${\cal O}_{\cal P}^{\nu}$ (${\cal O}_{\cal S}^{\nu})$\cite{MoRo,BMR} associated to the corresponding phase  of the RG analysis (l.h.s. of the Table).\\

The four partly gapped phases  appear to be SPT phases in one channel, the 
other channel remaining gapless. The further four fully gapped phases appear in 
correspondence of coexisting SPT phases in the charge and spin channels. Their 
nature can be better understood by noticing that within bosonization analysis the 
simoultaneous pinning of the two bosonic fields manifests in the non-zero value of an appropriate
local correlator\cite{Giam}, which is the hallmark of a spontaneous breaking
of a symmetry. The effect is a consequence of the limited number of independent bosonic fields in
1D. One coud argue that, as soon as the one-dimensionality of the
system is released, states with coexisting SPT phases and no spontaneous
symmetry breaking appear\cite{DMR}.

\section{Discussion and conclusions}

In this article we have analyzed the interaction induced topological properties of the 1D 
fermionic models with spin-charge separation, described in the  low energy limit by two decoupled
sine-Gordon models.
Specifically, we have shown that the rich phase diagram stemming from the RG-analysis of the sine-Gordon models --which gapped phases are Mott, density waves, and Haldane insulators, and Luther Emery liquids, characterized by different non-local orders (l.h.s. of Table~I)-- contains phases protected by P symmetry which are topologically non-trivial (column ~IV, r.h.s. of the Table). Importantly, we have proven that it is in one-to-one correspondence with the SPT phases classification based on the group cohomology theory (column ~V, r.h.s. of the Table).
The results represent a conceptual bridge between the two types of phase classification, and have relevant physical implications. On the one side, one can now identify in each (charge or spin) channel the topologically non-trivial phase among those stemming from RG analysis. Such non trivial phase, which is protected by time-reversal T and/or particle-hole  P symmetry, hosts edge modes with fractionalized charge/spin, degenerate for appropriate magnetization/filling values. On the other hand, our results provide a  physical characterization of the group cohomology classification, by associating to each topological phase a non-local order described by a non-vanishing expectation value of suitable and well identified operators.  As an application, we can for instance predict that ${\cal O}_{\cal S}^s$ is non-vanishing in the SPT phases discussed in \cite{KaCa,KeBe}.
 
The results,  based on the assumption of irrelevance of spin-charge coupling terms in the interaction, exploit the two continuous U(1) symmetries related to charge and spin-$z$ conservation and the discrete particle-hole symmetry P of the sine-Gordon model. In this respect, we point out that symmetries, which are crucial in the topological classification, are not necessarily fulfilled by lattice models  over the whole energy spectrum. As a matter of fact, very few 1D fermionic models were observed to host an interaction induced topologically non-trival phase~\cite{BMR,NJL,LEF}. However, the asymptotic properties of a model are well captured by the effective Hamiltonian at low energies, where  additional symmetries may arise. This is precisely the case for the P symmetry:  when considering the wide class of extended Hubbard  models, the P-symmetry is typically {\it not} fulfilled. However, because the asymptotic properties are described by decoupled sine-Gordon models, such symmetry does emerge in the low energy sector, enabling one to determine the topological properties of the existing phases.
For these reasons our result  extends  the topological characterization to a wide class 
of extended Hubbard systems, e.g. with correlated and/or pair hopping terms\cite{DoMo2}, spin-orbit coupling   \cite{KeBe,KaCa}, and 3- and 4-body couplings\cite{DoMo}.  

A remark can also be made about the role of  NLO parameters. 
In particular, we have shown that the nontrivial phase is characterized by a non-zero Haldane NLO, whereas
the trivial phase corresponds to the finiteness of the parity NLO.  Such correspondence suggests 
the possibility to identify nontrivial topological phases in microscopic lattice systems by the observation of a finite 
Haldane string value, both in the numerical investigation of regimes where weak coupling bosonization does not hold\cite{FMRB}, and in experimental simulations\cite{Eal,Hal}. 
Moreover, parity NLO could be useful to detect distinct trivial phases, whose presence was  
recently recognized in specific spin systems\cite{FPO}. The generalization of parity operators 
to ladder systems~\cite{DMR,FBM} suggests that distinct trivial phases could persist as long
as spin-charge separation does. In fact, this may be the case also
for some non-trivial phases \cite{YeWa}.

We point out another consequence of the one-to-one correspondence described by Table~I. 
The ordered phases characterized by spontaneous symmetry breaking  correspond
to the simultaneous occurrence of two SPT phases, one in the charge
and the other in the spin channel. In the context of SPT phases, the occurrence of spontaneus symmetry
breaking  was recently discussed in Ref.[\onlinecite{ZeWe}],
where it was noticed that such states  can be regarded to as  non-trivial
gapped quantum liquids, with a ground state degeneracy that becomes
unstable. 

Finally, we would like to mention   possible future developments of our results.
First, the other discrete symmetry characterizing $\mathcal{H}_{SG}$, i.e. time-reversal, has not been explicitly harnessed in our derivation. We thus expect that our results about the topological classification may be extended to cases where the low energy Hamiltonian contains further time-reversal breaking terms, as long as P symmetry holds.  
Secondly, when the  microscopic
Hamiltonian  also break the U(1) spin/charge symmetry \cite{GJPB,KaCa}, the continuum model should contain marginal terms also
in the dual fields $\theta_{\nu}$, so that further phases could modify
the present classification. Their characterization by means of other
NLO parameters is thus an interesting open issue.

\begin{acknowledgments}
A.M. acknowledges MITOR 2015 funds ``Nonlocal'' and the hospitality
of the Physics Department of MIT, where this work was conceived. As
well as useful discussions with Liang Fu, Senthil Todadri, Jonathan
Ruhman. F.D. also acknowledges financial support by the Italian FIRB
2012 project HybridNanoDev 976 (Grant No. RBFR1236VV).\end{acknowledgments}

\bibliographystyle{apsrev}
\bibliography{entanglement}

\begin{thebibliography}{10}

\bibitem{Hubb} J. Hubbard, Proc. Roy. Soc. (London), {\bf A276}, 238
(1963)

\bibitem{Giam}T. Giamarchi, {\it Quantum Physics in one Dimension},
(Oxford University Press, 2003)

\bibitem{GNT} A. O. Gogolin, A. A. Nersesyan, A. M. Tsvelik, {\it Bosonization and Strongly Correlated Systems} (Cambridge Univ. Press, 1998)


\bibitem{JaMH} J. Japaridze and E. M\"uller-Hartmann, Ann. Phys. (Leipzig) {\bf 3}, 163 (1994)

\bibitem{JaKa} G. I. Japaridze, A. P. Kampf, Phys. Rev. B {\bf 59}, 12822
(1999)
\bibitem{DoMo} F. Dolcini, and A. Montorsi, Phys. Rev. B {\bf 88}, 115115
(2013)

\bibitem{MoRo} A. Montorsi, and M. Roncaglia, Phys. Rev. Lett. {\bf 109}, 236404   (2012).

\bibitem{BMR} L. Barbiero, A. Montorsi, and M. Roncaglia, Phys. Rev.
B {\bf 88}, 035109 (2013) (2013)

\bibitem{KeBe} A. Kesselman, E. Berg, Phys. Rev. B {\bf 91}, 235309 (2015)

\bibitem{KaCa} N. Kainaris, and S.T. Carr, Phys. Rev. B {\bf 92}, 035139
(2015)

\bibitem{AZ} A. Altland, and M. R. Zirnbauer, Phys. Rev. B {\bf 55}, 1142
(1997)

\bibitem{SRFL} A.P. Schnyder, S. Ryu, A. Furusaki, A.W. Ludwig, Phys.
Rev. B {\bf 78}, 195125 (2008)

\bibitem{Niro} M. den Nijs, K. Rommelse, Phys. Rev. B {\bf 40}, 4709 (1989)

\bibitem{Eal} M. Endres et al., Science {\bf 334}, 200 (2011)

\bibitem{Wen} X.-G. Wen, Adv. Phys. {\bf 44}, 405 (1995)

\bibitem{NuOr} Z. Nussinov, G. Ortiz, Proc. Natl. Acad. Sci. {\bf 106}, 16944 (2009)

\bibitem{PTBO} F. Pollmann, A.M. Turner, E. Berg, and M. Oshikawa,
Phys. Rev. B \textbf{81}, 064439 (2010)

\bibitem{GuWe} Z.-C. Gu, X.-G. Wen, Phys. Rev. B {\bf 80}, 155131 (2009)

\bibitem{PBTO} F. Pollmann, E. Berg, A. M. Turner, M. Oshikawa, Phys.
Rev. B {\bf 85}, 075125 (2012)

\bibitem{CGW} X.Chen, Z.-C. Gu, and X.-G. Wen, Phys. Rev. B {\bf 83}, 035107 (2011); {\bf 84}, 235128
(2011)

\bibitem{CGLW} X. Chen, Z.-C. Gu, Z.-X. Liu, and X.-G. Wen, Science
{\bf 338}, 1604 (2012); 

\bibitem{CGLW2} X. Chen, Z.-C. Gu, Z.-X. Liu, and X.-G. Wen, Phys. Rev. B {\bf 87}, 155114 (2013)

\bibitem{WaWe} E. Wang, X.-G. Wen, Phys. Rev. Lett. 109, 096403 (2012)

\bibitem{Wen-1} B. Zeng, X. Chen, D.-L. Zhou, X.-G. Wen, preprint ArXiv:1508.02595

\bibitem{SPC} N. Schuch, D. P\'erez-Garc\'ia, and I. Cirac, Phys. Rev.
B {\bf 84}, 165139 (2011).

\bibitem{Fuji} Y. Fuji, Phys. Rev. B {\bf 93}, 104425 (2016)

\bibitem{AlAr} A. A. Aligia and L. Arrachea, Phys. Rev. B {\bf 60}, 15332 (1999)

\bibitem{Naka} M. Nakamura, Phys. Rev. B {\bf 61}, 16377 (2000)

{\bibitem{Fabr} M. Fabrizio, Phys. Rev. B {\bf 54}, 10054 (1996)}

\bibitem{DoMo1} F. Dolcini, and A. Montorsi, Nucl. Phys. {\bf B592}, 563 (2001) 

\bibitem{GJPB}V. Gritsev, G. Japaridze, M. Pletyukhov, and D. Baeriswyl,
Phys. Rev. Lett. {\bf 94}, 137207 (2005)

\bibitem{BDGA} E. Berg, E.G. Dalla Torre, T. Giamarchi, and E. Altman,
Phys. Rev. B {\bf 77}, 245119 (2008)

\bibitem{FK} L. Fidkowski, and A. Kitaev, Phys. Rev. B {\bf 83} 075103
(2011)

\bibitem{MFM} T. Morimoto, A. Furusaki, C. Mudry, Phys. Rev. B {\bf 92},
125104 (2015)

\bibitem{TPB}A.M. Turner, F. Pollmann, and E. Berg, Phys. Rev. B
{\bf 83}, 075702 (2011)

\bibitem{Faal} C. Fasth, A. Fuhrer, L. Samuelson, V. N. Golovach,
and D. Loss, Phys. Rev. Lett. {\bf 98}, 266801 (2007)

\bibitem{Citro} E. Orignac, R. Citro, M. Di Dio, S. De Palo, and
M.-L. Chiofalo, New J.  Phys. {\bf 18}, 055017 (2016)

\bibitem{AnRo} F. Anfuso and A. Rosch, Phys. Rev. B {\bf 75}, 144420 (2007)

\bibitem{MoPo} S. Moudgalya, and F. Pollmann, Phys. Rev. B {\bf 91}, 155128
(2015)

\bibitem{DMR}C. Degli Esposti Boschi, A. Montorsi, and M. Roncaglia,
Phys. Rev. B {\bf 94}, 085119 (2016)

\bibitem{NJL}S.-Q. Ning, H.-C. Jiang, and Z.-X. Liu, Phys. Rev. B
{\bf 91}, 241105(R) (2015)

\bibitem{LEF}F. Lange, S. Ejima, H. Fehske, Phys. Rev. B {\bf 92}, 041120
(2015)

\bibitem{DoMo2} F Dolcini, and A. Montorsi, Phys. Rev. B {\bf 65} 155105 (2002); {\it ibid.} {\bf 66}, 075112 (2002)

\bibitem{FMRB} S. Fazzini, A. Montorsi, M. Roncaglia, and L. Barbiero, preprint ArXiv:1607.05682 

\bibitem{Hal} T.A. Hilker, G. Salomon, F. Grusdt, A. Omran, M. Boll, E. Demler, I. Bloch, and C. Gross, preprint ArXiv:1702.00642

\bibitem{FPO}Y. Fuji, F. Pollmann, M. Oshikawa, Phys. Rev. Lett.
{\bf 114}, 177204 (2015)

\bibitem{FBM} S. Fazzini, F. Becca, and A. Montorsi, Phys. Rev. Lett. 118, 157102 (2017)

\bibitem{YeWa} P.Ye, and J. Wang, Phys. Rev. B {\bf 88}, 235109 (2013)

\bibitem{ZeWe} B. Zeng, and X.-G. Wen, Phys. Rev. B {\bf 91}, 125121 (2015)




\end{thebibliography}

\end{document}